\documentclass[prc,preprint,showpacs,showkeys,nofootinbib,tightenlines]{revtex4}

\usepackage{graphicx,color}  %
\usepackage{bm}  %

\newcommand{\beq}{\begin{equation}}
\newcommand{\eeq}{\end{equation}}
\newcommand{\bea}{\begin{eqnarray}}
\newcommand{\eea}{\end{eqnarray}}
\newcommand{\ben}{\begin{eqnarray*}}
\newcommand{\een}{\end{eqnarray*}}

\newcommand{\simle}{\hspace*{0.2em}\raisebox{0.5ex}{$<$}
     \hspace{-0.8em}\raisebox{-0.3em}{$\sim$}\hspace*{0.2em}}

\renewcommand{\vec}[1]{{\mathbf #1}}

\renewcommand{\Re}[1]{{{\rm Re\,} #1}} 
\renewcommand{\Im}[1]{{{\rm Im\,} #1}} 

\def\lixo#1{}

\begin{document}

\title{$\alpha\alpha$ Scattering in Halo Effective Field Theory}

\author{R. Higa}\email{higa@itkp.uni-bonn.de}
\author{H.-W. Hammer}\email{hammer@itkp.uni-bonn.de}
\affiliation{Helmholtz-Institut f\"ur Strahlen- und Kernphysik, 
Universit\"at Bonn, 53115 Bonn, Germany}
\author{U. van Kolck}\email{vankolck@physics.arizona.edu}
\affiliation{Department of Physics, University of Arizona,
          Tucson, AZ\ 85721, USA}
\affiliation{Kernfysisch Versneller Instituut, Rijksuniversiteit Groningen, 
Zernikelaan 25, 9747 AA Groningen, The Netherlands}
\affiliation{Instituto de F\'{\i}sica Te\'{o}rica, 
Universidade Estadual Paulista,
Rua Pamplona 145, 01405-900 S\~{a}o Paulo, SP, Brazil}

\today

\begin{abstract}
We study the two-alpha-particle ($\alpha\alpha$) system in an 
Effective Field Theory (EFT) for halo-like systems. 
We propose a power counting that incorporates
the subtle interplay of strong and electromagnetic forces
leading to a narrow resonance at an energy of about 0.1 MeV. 
We investigate the EFT expansion in detail, and compare its results 
with existing low-energy $\alpha\alpha$ phase shifts and previously 
determined effective-range parameters. 
Good description of the data is obtained with a surprising amount 
of fine-tuning. 
This scenario can be viewed as an expansion around the limit where, 
when electromagnetic interactions are turned off, the $^8$Be  ground 
state is at threshold and exhibits conformal invariance. 
We also discuss possible extensions to systems with more than two 
alpha particles.
\end{abstract}

\smallskip
\pacs{21.45.-v, 21.60.Gx}
\keywords{Effective field theory, nuclear clusters}
\maketitle  

\section{Introduction}
\label{sec:intro}

Nucleons in light nuclei have typical momenta that are small
compared to the characteristic QCD scale of 1 GeV.
At these low momenta, QCD can conveniently be
represented by a hadronic theory containing
all possible interactions consistent with the QCD symmetries.
Effective Field Theory (EFT) provides a controlled framework for 
exploiting the separation of scales in nuclei.
It is crucial to formulate a power counting that
justifies a systematic 
truncation of the Lagrangian leading to observables
with the desired accuracy.
Nuclei offer a non-trivial challenge because 
one wants such a perturbative expansion
in addition to the non-perturbative treatment of certain
leading operators, which is required by the existence of 
shallow bound states. By now, mainly 
few-body systems have been studied within EFT, and,
while much remains to be understood, many successes have been achieved 
\cite{Bedaque:2002mn,Braaten:2004rn}.

Similar to other approaches,
the extension of EFTs to larger nuclei faces computational
challenges \cite{Stetcu:2006ey,Borasoy:2006qn}.
As a first step in this extension, we specialized
to very low energies where clusters of nucleons 
behave coherently \cite{BHvK1,BHvK2,BRvK}.
Even though many interesting issues of 
nuclear structure are by-passed, we can describe anomalously 
shallow (``halo'' or ``cluster'') nuclei and some reactions of 
astrophysical interest.
Since they are strongly bound, 
alpha particles play a central role in this
framework. Many nuclear states 
have energies close to thresholds for break-up into 
alpha particles and nucleons, the most famous being
the excited (``Hoyle'') state of $^{12}$C near the 
triple-alpha ($3\alpha$) threshold.
These states should
be describable within the halo/cluster EFT, which is formulated
with contact interactions among nucleon ($N$) and alpha-particle ($\alpha$)
fields. Together with the $N\alpha$ interaction,  
the $\alpha\alpha$ interaction is an important input for such calculations.
While we have studied the $N\alpha$ interaction elsewhere
through both neutron-alpha ($n\alpha$) \cite{BHvK1,BHvK2} and 
proton-alpha ($p\alpha$) \cite{BRvK} scattering, we 
focus here on $\alpha\alpha$ scattering. 
Consideration of this system is required 
before tackling other states with 
two or more alpha particles, such as $^9$Be and $^{12}$C.

The internal alpha-particle dynamics is characterized by an intrinsic momentum
scale $M_{hi}$ associated with the binding mechanism.
A naive guess is that this scale is set by the pion mass $m_\pi\simeq 140$ MeV.
The $\alpha\alpha$ interaction consists of the long-range photon exchange
and short-range strong interactions. At low energies, the latter can
be represented by contact interactions.
The central issue is the relative importance of these contributions.
The Coulomb interaction is non-perturbative for momenta smaller than
around $k_C=Z_1 Z_2 \alpha_{em} \mu$, where 
$\alpha_{em}=e^2/4\pi$ is the fine-structure constant,
$\mu$ the reduced mass of the 
system, and $Z_i$, $i=1,2$, the electromagnetic charge of the particles.
Here $\mu=m_\alpha/2$ and $Z_i=Z_\alpha=2$ in terms of the
mass and charge of the alpha particle, respectively, so
$k_C\approx 60$~MeV.
At momenta much below 100 MeV, the deviation from pure-Coulomb
$\alpha\alpha$ scattering is dominated 
by the $S$ wave \cite{HT56,RPR56,TS63,JPM60,T66,R67,AAA69,Til04}.
The large near-threshold $S$-wave phase shift has been interpreted  
as resulting from a ($J^\pi$, $I$) = ($0^+$, 0) state 
\cite{Jon53,Ben68,Wue92,AAA69,Til04} at an energy 
$E_R\simeq 0.1$~MeV 
above threshold
in the center-of-mass frame, with a tiny width $\Gamma_R\simeq 6$ eV.
The momentum corresponding to this ${}^8$Be state sets a smaller 
scale $M_{lo}\sim k_R= \sqrt{2\mu E_R}\approx 20$ MeV, 
which must arise from the larger underlying scale $M_{hi}$ 
by a fine-tuning
of the parameters of the underlying theory. 

In the halo EFT, our goal is
not to explain the mechanism of this fine-tuning, but instead to 
exploit its existence in order to describe $\alpha$-cluster systems
at low energies. 
We seek a description of these systems in an expansion
in powers of the small ratio $M_{lo}/M_{hi}$.
Power counting is dependent on how the various 
parameters scale with $M_{lo}$ and $M_{hi}$.
The physics of the low-energy $S$ state is conveniently discussed
in the language of a dimeron field \cite{kaplan} 
with the quantum numbers of the low-energy composite state.
This field is characterized in leading order by 
a fine-tuned mass $\Delta$ and 
a non-derivative
coupling $g$ to the $\alpha\alpha$ state.
In subleading orders more complicated couplings appear.

It is not immediately obvious how the fine-tuned mass $\Delta$
relates to $M_{lo}$. 
The simplest assumption is
$\Delta \sim M_{lo}M_{hi}/\mu$ \cite{aleph}. 
In the absence of Coulomb interactions, this reproduces the 
leading term in the effective-range expansion, and one has a shallow
real or virtual bound state with a typical momentum $k_B\sim M_{lo}$. 
Strong interactions are non-perturbative for momenta of order
$k_B$ and larger.
Higher-order terms in the effective-range expansion
appear
as subleading corrections. 
This scenario is appropriate for $S$-wave $NN$ scattering
at momenta below $M_{hi} \sim m_\pi$ \cite{gautam,pole}.
For $pp$ scattering, $k_B\approx 8$ MeV
and $k_C\approx 4$ MeV.
The Coulomb interaction can be included
non-perturbatively in a straightforward way \cite{KR00},
providing calculable contributions plus a renormalization
of $\Delta/g^2$. 

The situation in $\alpha\alpha$ scattering is somewhat different.
The extremely 
low energy of the $S$-wave resonance
suggests that a smaller $\Delta$ might be necessary.
An alternative fine-tuning assumes thus
that $\Delta \sim M^2_{lo}/\mu$. 
Such scaling has already appeared in $P$-wave
$N\alpha$ scattering \cite{BHvK1,BHvK2},
and
has striking
consequences in $S$-wave $\alpha\alpha$ scattering.
In the absence of the Coulomb interaction,
the leading contribution 
for momenta $k\sim M_{lo}$
comes entirely from the unitarity term $-ik$
in the inverse amplitude. 
To this order, 
the ${}^8$Be system would be at the so-called
unitary limit, exhibiting conformal 
invariance \cite{msw,3bozos},
and ${}^{12}$C would acquire an exact Efimov spectrum
\cite{efimov,3bozos,Braaten:2004rn}.
As we will discuss later,
this exact pattern is modified by the Coulomb interaction, 
which breaks scale invariance
and thus moves the ${}^8$Be ground state away from threshold.
We will see that in leading order we can describe the
ground state, and higher-energy, scattering data
can be systematically accounted for in higher orders.
At this point, we simply fit the EFT parameters to data.
In the future, many-body methods \cite{Stetcu:2006ey,Borasoy:2006qn}
should allow the calculation of these parameters from
the underlying EFTs, which in turn will have their parameters
obtained from lattice QCD \cite{Beane:2006mx}.

To the extent that the halo/cluster EFT is built on fields
for clusters, it is closely related to phenomenological 
few-body cluster models
\cite{AAA69,clusterreview}.
The latter have had many applications,
ranging from the general analysis of halo structures
\cite{tobiasetal} 
to the successful confrontation with data for
specific processes, such as decays of $^{12}$C resonances
\cite{fedorovetal}.
The emphasis here is instead on a systematic expansion of the
most general interactions allowed by the symmetries of QCD.

In this paper, we study the alternative expansion scenario and perform a 
detailed comparison to phase-shift data for the $\alpha\alpha$ system. 
As we will see, among the many findings our new method entails are a new 
expansion for the $\alpha\alpha$ amplitude around the resonance, which 
differs from the effective-range expansion; the existence of inconsistencies 
between recent resonance measurements and old scattering data; a more 
precise determination of low-energy parameters; and an extraordinary amount 
of fine-tuning. All these findings could have implication in the study of 
other alpha-cluster systems. 
The paper is organized as follows.
In Sec.~\ref{sec:NPC}, we briefly discuss the EFT formulation of the 
two-body system with Coulomb interactions,
with some details 
relegated to App. \ref{app:CEFT}.
The proposed power counting is 
discussed in Sec.~\ref{sec:PC}. In Sec.~\ref{sec:fits},
we present results of our fits to the $0^+$ resonance parameters and 
$\alpha\alpha$ phase shifts, and discuss the required fine-tunings. 
Our conclusions are presented in Sec.~\ref{sec:conc}.

\section{EFT with Coulomb Interactions}
\label{sec:NPC}

We start with a summary of some basic ideas in 
halo EFT \cite{BHvK1,BHvK2,BRvK},
extended to include the main equations needed 
to deal with Coulomb interactions. 
These equations are a straightforward 
generalization of Kong and Ravndal's formalism~\cite{KR00} to include 
dimeron fields\footnote{A similar extension was recently considered 
by Ando and collaborators \cite{Ando07}.}. More details can be found in 
App.~\ref{app:CEFT}. In 
Sect.~\ref{sec:PC} we discuss the significant differences in scales 
between alpha-cluster systems and the $pp$ system considered by 
Kong and Ravndal~\cite{KR00}.
%

We consider here the scattering of two alpha
particles of charge $Z_{\alpha}=2$ and reduced 
mass $\mu=m_{\alpha}/2$, at a center-of-mass (CM) energy 
$E=k^2/2\mu$.
We want to build an EFT that provides a controlled expansion for
observables at momenta around the $0^+$ resonance,
$k\sim k_R=\sqrt{2\mu E_R}\approx 20$~MeV~$\sim M_{lo} \ll M_{hi}$.
The energy of the resonance, $E_R\simeq 0.1$ MeV,
is much smaller than either the alpha-particle
excitation energy $E^{*}_{\alpha}\simeq 20$~MeV
or any energy set by pion exchange, such as two-pion exchange
between alpha particles or one-pion exchange among nucleons,
$(2m_\pi)^2/m_\alpha \sim m_\pi^2/m_N  \approx 20$~MeV.
Thus, 
we expect an alpha particle to behave in a first approximation
as a rigid entity (``core'' or ``cluster''), which we represent by 
a scalar-isoscalar field $\phi$.
The smaller effects from deformations and other
core-structure properties are accounted for in a
derivative expansion.
The breakdown momentum scale $M_{hi}$ of this EFT will be set by the
lowest-energy 
degrees of freedom that are not incorporated explicitly in
the Lagrangian. Since these include the nucleons
within the alpha particle ---which can be resolved with 
a momentum $\sqrt{m_N E^{*}_{\alpha}}\approx 140$ MeV---
and the pions ---which can be resolved 
with momenta of the order of the pion mass $m_\pi\simeq 140$ MeV---
a reasonable estimate is 
$M_{hi}\sim \sqrt{m_N E^{*}_{\alpha}} \sim m_\pi \simeq 140$ MeV.
The EFT provides an expansion of observables in powers of the ratio
between the low-energy scales
$k$ and $M_{lo}$ and the high-energy scale $M_{hi}$.
\footnote{Recently it has been suggested \cite{ruizarriola} 
that chiral symmetry is important for the properties of
the $^{8}$Be ground state. This argument is based on an extension of the
halo EFT to include explicit pion fields. 
In fact, it had been pointed out before \cite{renato} that
the long-range interaction of alpha particles is related to two-pion exchange, 
but in the context of nucleon scattering on an alpha particle 
built out of four nucleons.
Since the pion mass is not much smaller than the typical nucleon momentum 
in the alpha particle, we see no rationale 
(for realistic values of the quark masses)
for an expansion in powers of $m_\pi/M_{hi}$ 
in a theory with ``elementary'' alpha particles.}
Around the resonance $k\sim k_R$, 
we expect an expansion parameter of the order of $1/7$. 
As the energy increases the expansion deteriorates
and one should not expect the EFT to be applicable to laboratory 
(LAB) energies much above $E_{LAB}=k^2/\mu \sim 3$~MeV,
a conservative estimate that corresponds to a CM momentum 
of about $70$~MeV. 

Since at low energies only $S$ waves contribute significantly
to $\alpha\alpha$ scattering beyond pure Coulomb scattering,
we introduce an auxiliary scalar dimeron field $d$ 
with ``residual mass'' $\Delta$, which couples to 
two $\alpha$ fields.
(Observables are independent of the choice of fields.)
We thus start with the following strong-interaction
effective Lagrangian:
\begin{eqnarray}
{\cal L}&=&
\phi^{\dagger}\Bigg[i\partial_0+\frac{\vec\nabla^2}{2m_{\alpha}}\Bigg]\phi
+\sigma\,d^{\dagger}\Bigg[i\partial_0+\frac{\vec\nabla^2}{4m_{\alpha}}
-\Delta\Bigg]d+g\,\Big[d^{\dagger}\phi\phi+(\phi\phi)^{\dagger}d\Big]
+\ldots\,,
\label{eq:LOLag}
\end{eqnarray}
where  $\sigma=\pm 1$ and $g$ is a coupling constant.
The ``$\ldots$'' represent interactions with higher derivatives. 
The sign $\sigma$ is used to match the sign of the 
effective range $r_0$. The existence of this sign merely reflects the 
non-physical, auxiliary character of the dimeron field~\cite{kaplan,aleph}. 
In momentum space, associated with
the displayed bilinear terms are the $\alpha$ propagator 
\begin{equation}
i\,S_{\alpha}(q_0;{\mbox{\boldmath $q$}})=\frac{i}
{q_0-{\mbox{\boldmath $q$}}^2/2m_{\alpha}+i\epsilon}\,,
\label{eq:alprop}
\end{equation}
and the dimeron propagator 
\begin{equation}
i\,D_{d}(q_0;{\mbox{\boldmath $q$}})=\frac{i\,\sigma}
{q_0-{\mbox{\boldmath $q$}}^2/4m_\alpha-\Delta+i\epsilon}\,.
\label{eq:dprop}
\end{equation}
Other terms in the Lagrangian (\ref{eq:LOLag}) can be treated as
insertions of interaction vertices in Feynman diagrams or corrections 
to the 
propagators (\ref{eq:alprop}) and (\ref{eq:dprop}).

Different assumptions about the dependence of $\Delta$ on $M_{lo}$
translate into different relative weights for the terms in the denominator
of Eq. (\ref{eq:dprop}), where 
$|\mbox{\boldmath $q$}|\sim\sqrt{2\mu q_0}= {\cal O}(M_{lo})$.
For example, in the present scenario all terms are of the same size, while 
in the $NN$ case the 
energy and kinetic terms are of subleading order and the dimeron propagator 
reduces in leading order to $i\,D_{d}\to -i\sigma/(\Delta-i\epsilon)$.
(The latter form gives rise in the amplitude not to a resonance but to a 
bound-state pole~\cite{aleph} at a momentum $k_B=i/a_0$, where 
$a_0$ is given by Eq.~(\ref{eq:ardef}).)
In this section we illustrate the use of the Lagrangian (\ref{eq:LOLag}) 
together with electromagnetic interactions to calculate $\alpha\alpha$ 
scattering amplitudes. We postpone a discussion of the relative importance 
of various terms until the next section. 

Electromagnetic interactions are introduced into the effective Lagrangian
(\ref{eq:LOLag}) in the standard way, that is, by 
both changing the derivatives into gauge-covariant ones,
and introducing gauge-invariant interactions generated by the
electromagnetic field strength. 
For practical calculations one usually works in a fixed gauge. 
The Coulomb gauge is very convenient, since it allows a clear separation 
between Coulomb and transverse photons. 
The former provides the leading electromagnetic interaction and is 
driven by the Sommerfeld parameter 
\begin{equation}
\eta(k)\equiv \frac{Z_{\alpha}^2\alpha_{em}\mu}{k}=\frac{k_C}{k} \, , 
\label{eq:etapar}
\end{equation}
with 
$\alpha_{em}\equiv e^2/4\pi$ the fine-structure constant 
and $k_C$ the inverse of the ``Bohr radius'' of the $\alpha\alpha$ system. 
This parameter is enhanced by the presence of 
$\mu$ in the numerator, which, as we are going to see,
makes $\eta$ large around the resonance and requires a 
non-perturbative resummation of Coulomb photons. 
On the other hand, transverse photons are suppressed 
\cite{Bedaque:2002mn,GLR07} by two powers of 
$M_{lo}/\mu$,
and when in loops by extra powers of $\alpha_{em}$.
Since numerically $\mu\sim 2 M_{hi}^2/M_{lo}$, 
these interactions contribute to orders 
---where they can be accounted
for in perturbation theory---
beyond the precision we are working with.
They can become significant only at energies comparable to the excitation 
energy of the $\alpha$ core.

The scattering amplitude for two particles interacting via Coulomb plus 
a short-range interaction is given by \cite{GW}
(for more details, see App. \ref{app:CEFT}):
\begin{equation}
T=T_C+T_{CS}\,,
\end{equation}
where $T_C$ and $T_{CS}$ are the pure-Coulomb and the 
Coulomb-distorted short-range amplitudes, respectively. 
Considering only $S$-wave interactions, the Coulomb-distorted short-range 
amplitude is parametrized in terms of the ``Coulomb-corrected'' phase 
shift $\delta^c_0$ as 
\begin{equation}
T_{CS}=-\frac{2\pi}{\mu}\,\frac{e^{2i\sigma_0}}{k(\cot\delta^c_0-i)}
=-\frac{2\pi}{\mu}\,\frac{C_{\eta}^2\,e^{2i\sigma_0}}
{2k_C \left[K(\eta)-H(\eta)\right]}\,,
\label{eq:TCStext}
\end{equation}
with the pure-Coulomb phase shift $\sigma_0$ given by Eq.~(\ref{eq:sigma_l}) 
in the Appendix and the Sommerfeld factor 
\begin{equation}
C_{\eta}^2=\frac{2\pi\eta}{e^{2\pi\eta}-1}
\,.
\end{equation}
The $H$ function is given by Eq.~(\ref{eq:Hdef}). For real $\eta$ 
it can be expressed as 
\begin{equation}
H(\eta)
=\Re[\psi(1+i\eta)]-\ln\eta+
\frac{i}{2\eta}C_{\eta}^2
\label{eq:H}
\end{equation}
\noindent 
in terms of the digamma function $\psi(z)=(d/dz)\ln\Gamma(z)$.
The other (real) term in the denominator of Eq. (\ref{eq:TCStext}) is 
the Landau-Smorodinsky $K$ function~\cite{LS44}, 
\begin{equation}
K(\eta)\equiv \frac{C_{\eta}^2}{2\eta}\,(\cot\delta^c_0-i)+H(\eta)\,.
\label{eq:Kfunc}
\end{equation}
At low energies it reduces to the effective-range expansion (ERE) 
in the presence of Coulomb interactions, 
\begin{equation}
K(\eta)= \frac{1}{2k_C}\left[-\frac{1}{a_0}+\frac{r_0}{2}k^2
-\frac{{\cal P}_0}{4}k^4+\dots\right]
\,,
\label{eq:Kapprox}
\end{equation}
where $a_0$, $r_0$, ${\cal P}_0$, $\ldots$, are the scattering length,
effective range, shape, $\ldots$, parameters.

As shown in App.~\ref{app:CEFT}, the calculation using 
the EFT Lagrangian (\ref{eq:LOLag}) leads to 
\begin{eqnarray}
\label{eq:TCS}
T_{CS}=-\frac{2\pi}{\mu}\,C_{\eta}^2\,e^{2i\sigma_0}\left[
\sigma\frac{2\pi\Delta^{(R)}}{\mu g^2}-\sigma\frac{\pi}{\mu^2 g^2}\,
k^2-2k_C H(\eta)\right]^{-1}+\dots\,,
\label{eq:TCSres}
\end{eqnarray}
where $\Delta^{(R)}$ is the Coulomb-renormalized mass parameter of the 
EFT Lagrangian, 
\begin{equation}
\Delta^{(R)}=
\Delta (\kappa)
-\sigma\frac{\mu g^2}{2\pi}
\left\{
\frac{\kappa}{D-3}
+2k_C\left[\frac{1}{D-4}
-\ln\left(\frac{\sqrt{\pi}\kappa}{2k_C}\right)-1+\frac{3}{2}\,C_E\right]
\right\}\,,
\label{eq:renorm}
\end{equation}
with $\kappa$ the renormalization scale and $D$ the dimension of spacetime.
Equation (\ref{eq:TCSres}) is in the form of the Coulomb-modified ERE with 
scattering length and effective range given respectively by 
\begin{equation}
a_0= -\sigma\frac{\mu g^2}{2\pi \Delta^{(R)}}\,,
\qquad 
r_0= -\sigma \frac{2\pi}{\mu^2 g^2}\,.
\label{eq:ardef}
\end{equation}

From Eq.~(\ref{eq:TCS}) it is clear that the effect of a non-perturbative 
Coulomb dressing of the strong-interaction amplitude ---apart from 
multiplying the amplitude by $C_{\eta}^2e^{2i\sigma_0}$ and from 
renormalizing the short-range parameters--- is to effectively replace 
the unitarity term $-ik$ by $-2k_C H(\eta)$.
In order to estimate the relative sizes of the various contributions
to Eq. (\ref{eq:TCSres}), we turn now to a discussion of power counting.
As we are going to see, at each order the EFT in the two-body system 
is equivalent to a truncation of the Coulomb-modified ERE.

\section{Power Counting} 
\label{sec:PC}

In this section we elaborate on the proposed power-counting scenario mentioned 
in the Introduction. We also give the $\alpha\alpha$ amplitudes that
are used in the next section for a comparison to phase-shift data.

In the present strongly-tuned scenario one considers 
$\Delta \sim M_{lo}^2/\mu$ and $g^2/2\pi \sim M_{hi}/\mu^2$, with other 
couplings scaling with $M_{hi}$ according to naive dimensional analysis. 
In this case the contribution of the bare dimeron propagator 
to the scattering amplitude at momentum $k\sim M_{lo}$ comes not only 
from the dimeron mass, but also from its kinetic term.
The simplest strong-interaction contribution is a bare dimeron 
propagator attached to initial and final external legs, which 
(as seen by multiplying Eq.~(\ref{eq:dprop}) by $g^2$)
is of ${\cal O}(2\pi M_{hi}/\mu M_{lo}^2)$. 
A non-Coulomb bubble diagram times another 
dimeron propagator brings an extra factor of
${\cal O}(\mu g^2 k/2\pi \Delta)={\cal O}(M_{hi}/M_{lo})$.
As a consequence, the bubble-chain
resummation is necessary. 
The resulting denominator acquires the form 
$(-1/a_0+r_0k^2/2-ik)$. Interesting in this power counting is the fact 
that at momenta $k\sim M_{lo}$
the first two terms are suppressed by $M_{lo}/M_{hi}$ compared to 
the last one. Therefore, all that is left at LO if the Coulomb
interaction is turned off
is the unitarity 
term $1/(-ik)$. 
In this limit, 
the ${}^8$Be system exhibits conformal 
invariance \cite{msw,3bozos} and the corresponding three-body system, 
${}^{12}$C, 
acquires an exact Efimov spectrum \cite{efimov,3bozos,Braaten:2004rn} . 
This scenario is a possible realization of the unitary limit.

This picture is significantly
modified when the Coulomb interaction is turned on.
In a non-relativistic system an $1/r$ potential breaks 
scale invariance and introduces the scale $k_C$
in the propagation of two charged particles.
As we have seen in Eq. (\ref{eq:TCS}),
the unitarity term is modified.
The balance between strong-interaction terms and Coulomb-modified
propagation depends on both the intrinsic strong-interaction scale
and $k_C$.
We note that the scales are very different in $\alpha\alpha$ than in $pp$
scattering.
While for the latter $k_C/k_B\sim 1/2$,
for the $\alpha\alpha$ system, 
$k_C=\alpha_{em} Z_\alpha^2 m_\alpha/2 \sim 60$~MeV 
and $k_R=\sqrt{m_\alpha E_R}\sim 20$~MeV, so $k_C/k_R\sim 3$.
For momenta $k\sim k_R$, we are therefore in the deep non-perturbative 
Coulomb region.\footnote{This fact suggests that one might be able to 
develop a perturbation scheme in powers of 
$k_R/k_C=2\sqrt{E_R/m_\alpha}/\alpha_{em} Z_\alpha^2$.}
This case corresponds to large values of $\eta$, where the function 
$2k_C H(\eta)$ is significantly different from 
the usual unitarity term $ik$. 
Using Stirling's series \cite{abramo}, 
\begin{equation}
\ln\Gamma(1+z)= \frac{1}{2}\,\ln 2\pi+\left(z+\frac{1}{2}\right)\ln z
-z+\frac{1}{12z}-\frac{1}{360z^3}+\cdots\,,
\end{equation}
in Eq. (\ref{eq:H}) gives
\begin{equation}
H(\eta)
= \frac{1}{12\eta^2}+\frac{1}{120\eta^4}+\cdots
+\frac{i\pi}{e^{2\pi\eta}-1}\,.
\label{eq:Happrox}
\end{equation}
The unitarity term 
is thus replaced by $2k_C H(\eta)\sim k^2/6k_C$. 
This term is now a factor
$k/6k_C$ smaller in magnitude 
than the unitarity term in the absence of Coulomb,
and comparable to the effective-range term coming from
the dimeron kinetic term.
This can be captured automatically if we take $3k_C \sim M_{hi}$,
as it appears to be the case numerically.

If all higher-order 
parameters are natural, in the $S$ wave
the shape-parameter term ${\cal P}_0 k^4={\cal O} (M_{lo}^4/M_{hi}^3)$ 
is down by $(M_{lo}/M_{hi})^2$ compared to the effective-range term
$r_0k^2={\cal O} (M_{lo}^2/M_{hi})$, 
and so should be each of the successive terms. 
On the other hand, 
the $D$-wave scattering-``length'' term 
$a_2 k^4={\cal O} (M_{lo}^4/M_{hi}^5)$ is down by $(M_{lo}/M_{hi})^6$
compared to the $S$-wave scattering length $a_0={\cal O} (M_{hi}/M_{lo}^2)$, 
and further energy dependence and higher angular momenta 
bring in further powers of $(M_{lo}/M_{hi})^2$. 

In this power counting, the 
Coulomb-distorted short-range amplitude $T_{CS}$ is given 
in LO  by Eq. (\ref{eq:TCSres}).
Including corrections in perturbation theory,
we obtain up to NLO:
\begin{eqnarray}
T_{CS}&=&-\frac{2\pi}{\mu}\,C_{\eta}^2\,e^{2i\sigma_0}\,\Bigg[
\frac{1}{-1/a_0\!+\!k^2\,r_0/2\!-\!2k_C H} 
+
\frac{{\cal P}_0}{4}
\frac{k^4}{(-1/a_0\!+\!k^2\,r_0/2\!-\!2k_C H)^2} 
\Bigg]\,,
\label{eq:ampsc2}
\end{eqnarray}
where ${\cal P}_0$ is given by a higher-derivative term in 
the Lagrangian (\ref{eq:LOLag}).
This corresponds to an expansion of the ERE formula, 
\begin{equation}
T_{CS}^{\rm (ERE)}=-\frac{2\pi}{\mu}\,\frac{C_{\eta}^2\,e^{2i\sigma_0}}
{-1/a_0\!+\!k^2\,r_0/2\!\!-k^4\,{\cal P}_0/4\!-\!2k_C H}\;,
\label{eq:ere}
\end{equation}
for small ${\cal P}_0$. 
To this order, all waves higher than $S$ are purely Coulombic.
Higher orders can be calculated similarly. 

Equation (\ref{eq:ampsc2}) holds for generic momenta $k\sim M_{lo}$.
However, it fails in the immediate proximity of $k_R$.
This situation is familiar from another application of the halo EFT to 
a resonance \cite{BHvK2}. The power counting works for $k\sim M_{lo}$ 
except in the narrow region $|k-k_R|= {\cal O}(M_{lo}^2/M_{hi})$ 
where the LO denominator nearly vanishes and a resummation of the NLO 
term, here associated with the shape parameter, is required. 
As one gets closer to the resonance momentum $k_R$, higher-order terms 
in the ERE are kinematically fine-tuned as well. 
This happens because the imaginary part of the denominator 
is exponentially suppressed by a factor $\exp(-2\pi\eta_R)\sim 10^{-8}$ 
and the real part gets arbitrarily small. (For 
$n\alpha$ scattering, multiple kinematical fine-tunings 
are prevented in the $P_{3/2}$ wave 
by the presence of an imaginary term $ik^3\sim M_{lo}^3$
\cite{BHvK1,BHvK2}.) 

This kinematical fine-tuning is not a conceptual problem. 
From the EFT point of view, each new fine-tuning is equivalent to 
reshuffling the series and redefining the pole position. Such a procedure 
works fine with a small number of kinematical fine-tunings, but is not 
practical in the present case. A better alternative is to perform an 
expansion around the resonance pole position.
The situation here is analogous to the $NN$ system, where we can choose
to expand the amplitude around the bound-state pole \cite{pole} rather 
than around zero energy.

A great simplification results from the fact that the resonance lies 
in the deep Coulomb regime, 
where Eq.~(\ref{eq:Happrox}) provides an accurate representation of 
$H$ up to the precision we are considering. 
The real terms shown in Eq.~(\ref{eq:Happrox})
can be seen as an expansion in powers
of $\sim (k/3k_C)^2={\cal O}(k^2/M_{hi}^2)$.
Of course in an asymptotic expansion
at some point the remaining terms can no longer be expanded; at that point
the remainder should be treated exactly.
In lowest orders, however, we can use the successive terms shown in 
Eq. (\ref{eq:Happrox}): 
numerically, the terms up to $\eta^{-4}$
work to better than 3\% for $E_{LAB}=3$~MeV.

The expansion (\ref{eq:Happrox}) makes the physics around the resonance
quite transparent.
Since the ``size'' of the resonance, $1/k_R$, is much larger
than the Bohr radius $1/k_C$, the Coulomb interaction is effectively 
short ranged, and the real part of $H$ is ERE-like.
This expansion matches well
with the expansion for the Landau-Smorodinski function,
which is in powers of $(k/M_{hi})^2$.
In $T_{CS}$,
not only the $k^2$ terms,
but also
higher-order terms  from strong and Coulomb interactions
have comparable sizes.
We can thus lump these terms together, 
defining 
\begin{equation}
\tilde{r}_0= r_0 -\frac{1}{3k_C}\, , 
\qquad \tilde{\cal P}_0={\cal P}_0\,+\,\frac{1}{15k_C^3}\,,
\label{eq:polepar2}
\end{equation}
and so on.

In this case we can rewrite up to NLO:
\begin{eqnarray}
T_{CS}&=&-\frac{2\pi}{\mu}\,\frac{C_{\eta}^2\,e^{2i\sigma_0}}
{-1/a_0+\tilde{r}_0k^2/2-\tilde{\cal P}_0 k^4/4-ikC_{\eta}^2}
\nonumber\\[1mm]&=&
-\frac{2\pi}{\mu}\,\frac{C_{\eta}^2\,e^{2i\sigma_0}}
{\tilde{r}_0(k^2-k_R^2)/2\!-\!\tilde{\cal P}_0(k^4\!-\!k_R^4)/4\!
-\!ikC_{\eta}^2}
\nonumber\\[1mm]&=&
-\frac{2\pi}{\mu}\,C_{\eta}^2\,e^{2i\sigma_0}\,\Bigg[
\underbrace{ 
\frac{1}{\tilde{r}_0(k^2-k_R^2)/2\!-\!ikC_{\eta}^2} }_{\rm LO\;\; term}
+\underbrace{ 
\frac{\tilde{\cal P}_0}{4}\,
\frac{(k^4\!-\!k_R^4)}
{(\tilde{r}_0(k^2-k_R^2)/2\!-\!ikC_{\eta}^2)^2}
}_{\rm NLO\;\; correction}+\ldots\,\Bigg],
\label{eq:poleTCS}
\end{eqnarray}
where
\begin{equation}
k_R^2 = \frac{2}{a_0 \tilde{r}_0}\,
\Bigg(\,\underbrace{1\raisebox{-11pt}{} }_{\rm LO\;\; term}
-\underbrace{\frac{\tilde{\cal P}_0}
{a_0 \tilde{r}_0^2}}_{\rm NLO\;\; correction}+\ldots\Bigg) \,.
\label{eq:polepos}
\end{equation}
From this expression one sees directly
that, indeed, $k_R\sim M_{lo}$,
with corrections of ${\cal O}(M_{lo}^2/M_{hi}^2)$.
Note that 
we keep the exact form of the imaginary term in Eq.~(\ref{eq:poleTCS}): 
even though it is negligible at $k\sim k_R$, 
it has an 
important exponential dependence on the energy 
responsible for keeping the phase shifts real in the elastic regime.

When $a_0<0$ and $\tilde{r}_0<0$, and thus $r_0< 1/3k_C$, 
we have $k_R^2>0$ and the two 
poles of Eq. (\ref{eq:poleTCS})
are located 
in the lower half of the complex-momentum plane very near the
real axis, as befits a very narrow resonance.
The amplitude $T_{CS}$ can be written in terms of the resonance energy
$E_R=k_R^2/2\mu$ and the resonance width $\Gamma(E)$ as 
\begin{equation}
T_{CS}=
\frac{2\pi e^{2i\sigma_0}}{\mu \sqrt{2\mu E}}\,\frac{\Gamma(E)/2}
{E-E_R+i \Gamma(E)/2}\, .
\label{eq:poleTCSwidth}
\end{equation}
One finds that
\begin{equation}
\Gamma(E)
=\Gamma(E_R) \frac{e^{2\pi k_C/k_R}-1}{e^{2\pi k_C/k}-1}
\Bigg[\,\underbrace{1\raisebox{-11pt}{} }_{\rm LO\;\; term}
-\underbrace{\frac{\mu^2 \tilde{\cal P}_0}{2\pi k_C} 
\left(e^{2\pi k_C/k_R}-1\right)
\frac{\Gamma(E_R)}{2} (E-E_R)\raisebox{-11pt}{} }_{\rm NLO\;\; correction}
+\ldots\Bigg]
\,,
\label{eq:widthstrongE}
\end{equation}
where
\begin{equation}
\Gamma(E_R)
=-\frac{4\pi k_C}{\mu \tilde{r}_0} \frac{1}{e^{2\pi k_C/k_R}-1}
\Bigg(\,\underbrace{1\raisebox{-11pt}{} }_{\rm LO\;\; term}
+\underbrace{\frac{\tilde{\cal P}_0k_R^2}{\tilde{r}_0}}_{\rm NLO\;\; correction}
+\ldots\Bigg)
\,.
\label{eq:widthstrong}
\end{equation}
The width is very small because of the large value of $2\pi k_C/k_R$
in the exponential. 

In the form of 
Eqs. (\ref{eq:poleTCSwidth}) and (\ref{eq:widthstrongE})
we can keep 
$E_R$
and $\Gamma(E_R)$ fixed at each order in the expansion.
Note that these equations
do not change to this order if one makes a different choice
---{\it e.g.}, $(k^2-k_R^2)^2$ instead of $k^4-k_R^4$---
for the form of the $\tilde{\cal P}_0$ term in Eq. (\ref{eq:poleTCS}).
Since at the resonance, $\delta_0^c(E_R)=\pi/2$ and
$[d\delta_0^c(E)/dE]_{E_R}=2/\Gamma(E_R)$,
the behavior of the phase shift around $E_R$ is fixed.
We turn now to a test of the expansion (\ref{eq:poleTCS})
mandated by our power counting.

\section{Comparison to Data}
\label{sec:fits}

In this section we briefly describe the experimental situation regarding 
$\alpha\alpha$ scattering at low energies. 
Our predictions 
are then compared against the available data and 
discussed in detail, with a particular emphasis on the scalings of the 
EFT parameters. 

Scattering data at low energies are not abundant.
At energies below $E_{LAB}=3$~MeV 
data were obtained, and a  phase-shift analysis performed, in Ref. \cite{HT56}.
All later references 
that we were able to find \cite{AAA69,Til04} start at higher energies.
For example, 
Ref. \cite{RPR56} 
covers the region $E_{LAB}=3$--6~MeV,
Ref. \cite{TS63} $E_{LAB}=3.8$--12~MeV,
and 
Ref. \cite{JPM60} $E_{LAB}=5$--9~MeV.
These data show that $\alpha\alpha$ scattering at 
$E_{LAB}\simle 6$~MeV is dominated by the $S$ wave,
thanks to the presence of a ($J^\pi$, $I$) = ($0^+$, 0) resonance immediately 
above threshold, 
identified as the ${}^8$Be ground state \cite{Til04}. 
Early determinations
of the $0^+$ resonance energy were performed in 
reactions like 
${}^{11}{\rm B}+p\to 2\alpha+\alpha$ 
\cite{Jon53}. 
Later measurements and careful analysis of the scattering of $^4$He
atoms off $^4$He$^+$ ions \cite{Ben68,Wue92}
improved the determination of 
the resonance energy and width to their currently accepted 
values, 
$E_{LAB}^R=184.15\pm0.07$~keV and $\Gamma_{LAB}^R=11.14\pm0.50$~eV. 
The resonance CM momentum is thus $k_R=\sqrt{\mu E_{LAB}^R}\approx 18.5$~MeV.
At 
$E_{LAB}\approx 6$~MeV the $D$-wave phase shift
crosses $\pi/2$, indicating the position of a broader resonance associated with
the first excited, ($2^+$, 0) state \cite{Til04}.
The $G$ wave does not become comparable to the $D$ wave until 
$E_{LAB}\sim 20$~MeV, interpreted as the region of an even broader 
($4^+$, 0) state \cite{Til04}.

The low-energy data can be studied with the ERE.
When used together with the 
$0^+$ resonance energy \cite{Jon53},
the 
data from Refs. \cite{HT56,RPR56} give
for these parameters \cite{RPR56}: 
$a_0=-1.76\times~10^{3}$~fm, $r_0=1.096$~fm, and 
${\cal P}_0=-1.654$~fm${}^{3}$. 
They provide a good description of the 
available phase shifts 
up to $E_{LAB}\approx 6$~MeV, which 
lends some credence to these numbers. 
However, it was pointed out \cite{T66} that without input
from the $0^+$ resonance 
width these parameters have large uncertainties.
Inclusion of both 
resonance energy and width from Ref. \cite{Ben68}
reduces these uncertainties considerably \cite{R67}:
$a_0=(-1.65\pm 0.17)\times~10^3$~fm, $r_0=(1.084\pm 0.011)$~fm, and 
${\cal P}_0=(-1.76\pm 0.22)$~fm${}^{3}$. 
Since the later, more precise 
data from Ref. \cite{Wue92}
are consistent with 
Ref. \cite{Ben68}, the 
ERE parameters from Ref. \cite{R67} 
can be seen as a reasonable parametrization 
of the existing data.

Here we use the phase shifts compiled in Table II of Ref.~\cite{AAA69}.
Since it is well-determined experimentally, and due to its relevance to 
the triple-alpha process, we use the $0^+$ state as an important constraint.
This is in line with the EFT approach, where lower-energy observables
have smaller theoretical errors.
It provides a relationship among our EFT parameters and, 
consequently, reduces the number of variables to be 
adjusted at each order in the power counting. 
Below, we also use the ERE from Ref. \cite{R67} for orientation,
and comment on the extremely large 
value of the scattering length $a_0$, which suggests a 
large amount of 
fine-tuning in the parameters of the underlying theory away from the 
naturalness assumption. 
In contrast, 
both $r_0^{-1}\sim 180$~MeV and ${\cal P}_0^{-1/3}\sim 170$~MeV 
correspond to natural scales comparable to the pion mass. 
At this level there is no evidence against our initial 
estimate of the expansion parameter $M_{lo}/M_{hi} \sim 1/7$.
Note that here we do 
not include dimeron fields for resonances beyond the ground state,
and therefore cannot go beyond 
the energy region where the $D$-wave resonance is significant.

In the power counting 
we are proposing for the $\alpha\alpha$ 
system, the amplitude $T_{CS}$ for generic momenta is given 
up to NLO by Eq. (\ref{eq:poleTCS}). As previously discussed, this 
expression combines the deep-non-perturbative Coulomb approximation 
(\ref{eq:Happrox}) for 
the function $H$ with the expansion around the resonance pole, 
which avoids the need for multiple kinematical fine-tunings. 
In LO, the two parameters $a_0$ and $\tilde{r}_0$ can be obtained 
from a fit to the resonance position and width. 
At NLO, scattering data are needed to determine $\tilde{\cal P}_0$.

\begin{figure}[t]
\includegraphics[height=6.0cm,width=8.0cm]{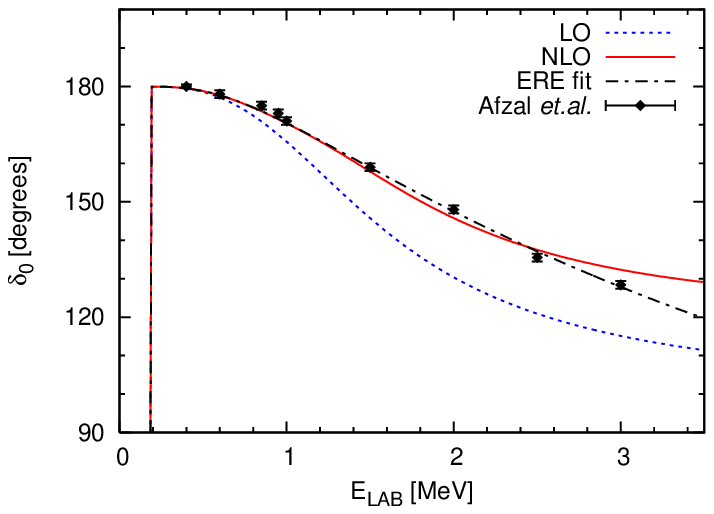}
\includegraphics[height=6.0cm,width=8.0cm]{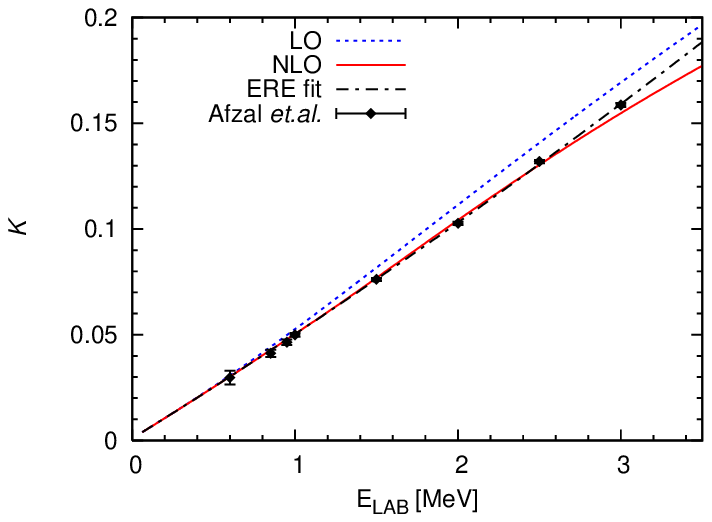}
\vspace*{-0pt}
\caption{Left panel: $S$-wave phase shift $\delta^c_0$ 
as a function of the laboratory energy $E_{LAB}$. 
The EFT results in LO and NLO 
are given by the (blue) dotted and (red) solid lines, respectively.
Our ERE fit using Eq.~(\ref{eq:ere}) is given by the dash-dotted line 
and the empirical phase shifts \cite{AAA69} as (black)
solid circles with error bars. 
Right panel: analogous results for the Landau-Smorodinsky $K$-function. 
}
\label{fig:pshift}
\end{figure}

Figure \ref{fig:pshift} shows the results of our fit to the available 
$S$-wave phase shifts below $E_{LAB}=3$~MeV, 
including the resonance position and width. 
The latter control the steep rise of the phase shift at very
low energies.
In the region above the resonance, where scattering data are shown,
the LO curve is a prediction, which is consistent with
the first few points but then moves away from the data.
The NLO curve has an extra parameter, which here was determined
from a global $\chi^2$-fit to scattering data shown.
As expected from a convergent expansion,
the description of the 
low-energy data improves with increasing order. 
At about 3 MeV and above, higher-order 
contributions are expected to be significant, as suggested by 
the discussion on the relevant scales and manifest in 
the growing difference between the NLO curve and both LO curve 
and data points. 
Also shown are results from a fit using the conventional ERE formula, 
Eq.~(\ref{eq:ere}), in order to stress the differences between this
and our EFT approach. 
The ERE formula includes some of the contributions
of higher order in the EFT. 

In Table~\ref{tab:erepar} we give the 
values of ERE parameters that we extract from the fits in 
Fig.~\ref{fig:pshift}, and compare them
with the values  \cite{R67} obtained from effective range theory.
At LO, $a_0$ and $r_0$ come out consistent with 
the values given in Ref. \cite{R67}, $r_0$ in particular. 
The changes due to NLO corrections, however, worsen this initial LO 
agreement. 
The reason for this deviation is most likely 
due to the calculation of the width constraint in Ref.~\cite{R67}. 
Its Eq.~(4) reads 
\begin{equation}
\frac{dh}{dk^2}(\eta_R)-\frac{1}{\mu\Gamma(E_R)}\,
\frac{\pi}{e^{2\pi k_C/k_R}-1}
=\frac{1}{4k_C}\left(r_0-{\cal P}_0\,k_R^2\right),
\end{equation}
where $h(\eta)\equiv \mbox{Re}[H(\eta)]$. 
Following 
this reference's procedure, we were able to reproduce 
the quoted value of the width (6.4 eV) only when 
$dh(\eta_R)/dk^2$ 
was approximated by $1/12k_C^2$ [see Eq.~(\ref{eq:Happrox})]. 
This amounts to ignoring the 
electromagnetic piece of $\tilde{\cal P}_0$ in Eq.~(\ref{eq:widthstrong}), 
which is inconsistent since the strong piece contributes at the same order. 
It explains why the value for $r_0$ in Ref. \cite{R67}
agrees quite well with ours at LO 
but disagrees at NLO. With this larger $\tilde r_0$, one can also 
understand why Ref. \cite{R67} obtains a smaller $a_0$ (see the following 
discussion on the scaling of $a_0$). 
When we repeated 
Ref. \cite{R67}'s procedure including the width constraint 
consistently, we obtained essentially the same values as in our EFT 
fits. This updated ERE fit is also shown in Table~\ref{tab:erepar}. 

\begin{table}[b]
\begin{center}
\begin{tabular} {|c||c|c|c|}
\hline
 & $a_0$ ($10^3$~fm) & $r_0$ (fm) & ${\cal P}_0$ (fm${}^{3}$) 
\\ \hline\hline
LO & $-1.80$ & 1.083 & --- \\ \hline
NLO & $-1.92\pm 0.09$ & $1.098\pm 0.005$ & $-1.46\pm 0.08$ \\ \hline
ERE (our fit) & $-1.92\pm 0.09$ & $1.099\pm 0.005$ & $-1.62\pm 0.08$ \\ \hline
ERE~\cite{R67} & $-1.65\pm 0.17$ & $1.084\pm 0.011$ & $-1.76\pm 0.22$ 
\\ \hline
\end{tabular}
\end{center}
\caption{ERE parameters 
extracted from EFT fits
in the first two orders, 
compared with values from two ERE fits, our own 
and Ref.~\cite{R67}'s.
\label{tab:erepar}}
\end{table}

In agreement with
our expectations, the effective range shows a natural size, 
$r_0\sim 1/(180\mbox{ MeV})$. 
The shape parameter 
also has a
natural size 
${\cal P}_0\sim 1/(170\mbox{ MeV})^3$ at NLO
which is in 
agreement with our {\it a priori} estimate.
The relative errors in $a_0$ and $\tilde r_0$ at LO are estimated to 
be of the order of the EFT parameter expansion, 
$M_{lo}/M_{hi}\sim 1/7\approx 15$\%. 
At NLO, they 
are essentially given 
by the precision of the most recent measurement of the resonance width 
\cite{Wue92}, which lies between 4--5\%.
The uncertainty in $\tilde{\cal P}_0$, given by the $\chi^2$-fit,
is of the same order. 
Note that the small relative error in $r_0$ compared to the one 
in $\tilde r_0$ is due to the former being an order of magnitude larger 
than the latter, as we discuss below.
The NLO errors found here are a factor 
of two smaller than the ones obtained by Ref.~\cite{R67}. 

One should stress that the accurate value of the resonance 
width, $\Gamma_R=5.57\pm 0.25$~eV, imposes tight constraints on 
our fits, through 
$a_0$ and $r_0$. 
There is a significant 
improvement in our NLO fit and overall agreement with data, 
but the theoretical 
curve is not able to cross the error bands of all scattering points 
below 3~MeV, as reflected in a $\chi^2$/datum 
$\simeq 4$.
In principle, a better agreement should be achieved by an N${}^2$LO 
calculation. However, that introduces an extra parameter that is mostly 
determined by the scattering data, and this agreement could mask any 
possible inconsistencies that the phase shifts might have with the 
resonance parameters. 
The high NLO $\chi^2$/datum suggests 
that the resonance width and the $S$-wave 
scattering data set are not compatible with each other or, at least, 
one of them has overestimated precision. As a test we performed fits 
to scattering data without the input from the resonance width, 
within our NLO EFT and the conventional ERE framework. In both cases, 
description of $S$-wave phase shifts is much better 
but the 
resonance width is underpredicted, 
$\Gamma_R=4.9\pm 0.6$~eV with ERE and $\Gamma_R=2.87\pm 0.23$~eV with EFT. 
The ERE result is still consistent with the measured $\Gamma_R$ 
thanks to its large 
error bar. In EFT, where lower-energy data have higher priority, 
the discrepancy is amplified. 
The problem is even more pronounced if the fit is performed using data up to 
2.5~MeV instead of 3~MeV: the results 
$\Gamma_R=4.2\pm 0.6$~eV with ERE and $\Gamma_R=2.93\pm 0.34$~eV with EFT
fall beyond the quoted experimental
error bars. Oddly, this tendency continues as one lowers the upper limit 
in the fit. 
Reanalyses of the existing low-energy data or even new measurements 
seem necessary to resolve this discrepancy. 

An astonishing 
feature of the $\alpha\alpha$ system is the very large 
magnitude of $a_0$, even if compared
to the large scattering length observed in the two-nucleon system. 
The latter is evidence of a fine-tuning in the QCD parameters,
which gives rise to an anomalously low momentum scale.
(This fine-tuning can be seen as the proximity
of the observed pion mass to the critical
value $\sim 200$ MeV where the $NN$ bound states have zero energy
\cite{critical,Beane:2006mx} and the triton displays the Efimov 
effect \cite{irlimcycl}.)
Certainly this fine-tuning scale propagates to heavier systems.
However, the enormous value of $a_0$ in the $\alpha\alpha$ system
is suggestive of a more dramatic fine-tuning, with electromagnetic 
interactions playing a crucial role. 

The parameters of the $0^+$ resonance are indeed quite surprising.
As we remarked, this resonance is associated with two poles of $T_{CS}$,
the momenta of which are much smaller in magnitude than
$k_C$.
As a consequence,
there is an exponential suppression of the width, which is evident
in Eq. (\ref{eq:widthstrong}). Yet, the width is somewhat
{\it large} given the position of the resonance.
It is remarkable that in order to fit both the resonance position
and width one
needs $\tilde{r}_0 \simeq -0.13$ fm, nearly a factor 10 smaller
in magnitude
than the Coulomb contribution to the 
energy-dependent term, 
$-1/3k_C\simeq -1.2$ fm. 
That is, although both
$r_0\sim 1$ fm and $1/3k_C\sim 1$ fm as expected from dimensional
analysis, they approximately cancel producing a balance
that is about 10\% of either. 
This balance would itself have natural size
if the width were a factor of
10 smaller than measured, or else if $k_R$ were 15\% larger. 
Instead, the existing cancellation
effectively leads to
either $|\tilde{r}_0|\sim 1/\mu$ or 
$|\tilde{r}_0|\sim M_{lo}/M_{hi}^2$. 

Since the position of the resonance determines to 
a first approximation $a_0\tilde{r}_0$, 
the seemingly accidental cancellation in $\tilde{r}_0$ translates into 
an exceptionally large $a_0$, 
which is effectively $|a_0|\sim \mu/M_{lo}^2$ or
$|a_0|\sim M_{hi}^2/M_{lo}^3$.
The Coulomb-modified parameter $\Delta^{(R)}$ is thus
suppressed by another
power of $M_{lo}/M_{hi}$ compared to
the assumed scaling 
when Coulomb interactions are turned off. 

The values of ERE parameters show that, as anticipated
by the power counting we proposed, 
the strong-interaction effective-range term is comparable
to both the scattering-length term and the
electromagnetic $H$. A power counting based on the scalings
appropriate for $NN$ scattering would instead miss the resonance
in LO unless one also counted $3k_C$ as $M_{lo}$,
which is a numerical stretch. Even so,
the width would be far off, since $1/3k_C\gg |\tilde{r}_0|$,
and convergence questionable.

Although the results presented in this section were based 
on Eq. (\ref{eq:poleTCS}), and thus assumed the validity of 
the expansion (\ref{eq:Happrox}),
we have checked that no qualitative changes arise if one bases
the EFT fits on Eq. (\ref{eq:ampsc2}) instead. In the latter case,
one is simply retaining at each order some terms that truly
belong to higher orders, at the cost of much less transparent 
analytical expressions.

\section{Conclusion}
\label{sec:conc}

In this paper, we have studied 
$\alpha\alpha$ interactions within the framework of halo/cluster EFT. 
From this perspective, the $\alpha\alpha$ system is 
most extraordinary.

EFT is model independent to the extent that includes all
interactions allowed by known symmetries.
It is equivalent in the $\alpha\alpha$ system to a truncation of the ERE.
The halo EFT can be viewed as a formulation of the ERE that
can be used for other systems
made of shallow-bound alpha particles and, together with results from
Refs. \cite{BHvK1,BHvK2,BRvK}, also nucleons. 

We have designed a power-counting scheme that seems 
realistic for 
the $\alpha\alpha$ system, but requires even more fine-tuning than 
the one in the $NN$ system. It assumes that the parameter $\Delta$ in 
the EFT Lagrangian scales as $M_{lo}^2/\mu$ in the
absence of the Coulomb interaction.
Interesting in this scenario is the fact that at LO
without Coulomb the ${}^8$Be system 
would exhibit conformal 
invariance and ${}^{12}$C would display
an exact Efimov spectrum. These exact features are broken by 
the Coulomb interaction but some remnants of this behavior are 
manifest in the experimental spectra,
such as the shallowness of the $0^+$ ${}^8$Be resonance.

We have incorporated two new ideas in the halo/cluster EFT.
The most effective way of implementing an EFT with resonances seems to 
be an expansion around the resonance pole, since it avoids the need for 
multiple kinematical fine-tunings and improves convergence. 
Our study was significantly simplified by treating the Coulomb function 
$H(\eta)$ 
in a power series expansion in $k/k_C$.
This simplification may be very relevant when treating interactions 
with several alpha-particle clusters. 

We made use of the 
precise measurements of the $0^+$ resonance properties 
\cite{Wue92}, together with the (rather old) existing $S$-wave scattering 
phase shifts \cite{HT56} in order to produce LO and NLO fits that
fix our ERE parameters. 
The result is in slight disagreement with Ref.~\cite{R67}, probably 
due to an 
approximation made in its calculation of the width. 
The uncertainties in our ERE parameters 
are smaller by 
a factor of two in comparison with this reference.
A systematic improvement is seen in the theoretical phase shifts,
and 
reasonably good overall description 
is obtained at NLO. 
Yet,
the rather high $\chi^2$ when the resonance width is fitted and
the deterioration of the calculated width as one
restricts the fit to lower-energy scattering data 
suggest 
that
either the resonance width or the $S$-wave phase shifts have 
overestimated precision. 

Despite the phenomenological success of the proposed power counting, 
the necessary cancellations between strong and Coulomb interactions,
which seem accidental, are puzzling.
In the absence of Coulomb interactions, the regulator-dependent
part of the bare dimeron mass $\Delta$ is ${\cal O}(M_{hi}^2/\mu)$.
As we can see from Eq. (\ref{eq:renorm}), the electromagnetic
$\kappa$-dependent pieces do not change this expectation,
since they are $\sim 2k_C \mu g^2/2\pi = {\cal O}(M_{hi}^2/\mu)$.
Therefore, barring cancellations between the finite pieces of 
$\Delta$ and loops, we would expect 
the renormalized mass $\Delta^{(R)}$ to have the same size,
which would set a scale for $|a_0|$ at
${\cal O}(1/M_{hi})\sim 1/2k_C \simeq 2$ fm. 
Yet, the observed resonance energy (together with a natural-sized
effective range $\tilde{r}_0$)
tells us that $\Delta^{(R)}={\cal O}(M_{lo}^2/\mu)$.
A similar scaling for $\Delta^{(R)}$ emerges in $n\alpha$ scattering
\cite{BHvK1,BHvK2}, but here we need a further cancellation between
the finite pieces in $\Delta$ and Coulomb loops.
This brings 
$|a_0|$ up by two orders of magnitude,
or more than
two inverse powers of
the expansion parameter $M_{lo}/M_{hi}\sim 1/7$, to 
${\cal O}(M_{hi}/M_{lo}^2)\simeq 100$ fm. 

Uncomfortable as this fine-tuning by a factor of nearly 100
in the energy-independent part
of the amplitude $T_{CS}$ might be, it is not the whole story.
As we have shown, both the resonance 
width and the higher-energy 
phase shifts require a fine-tuning also in the energy dependence
of the amplitude. This $\sim 90$\% additional cancellation
(together with the observed resonance energy)
further enhances $|a_0|$ by another factor of 10,
or about another inverse power of $M_{lo}/M_{hi}$,
leading effectively to ${\cal O}(M_{hi}^2/M_{lo}^3)\simeq 700$ fm,
which is indeed the order of magnitude we obtain in our fit. 
This fine-tuning of a factor of $\sim$ 1000 in $\alpha\alpha$
completely overshadows the fine-tuning of $\sim$ 10 
---from $m_\pi$ (or from the pion decay constant $f_\pi\simeq 92$ MeV)
down to $1/|a_0|\simeq 8$ MeV--- in the $NN$
$^1S_0$ channel.
It has important consequences: for example,
if the strong-interaction effective range
$r_0$ were just 15-20\% larger it would make the
ground state of $^8$Be bound, presumably with far-reaching
effects in nucleosynthesis. 

This context frames a fascinating picture
for the $\alpha\alpha$ system:
a nearly conformally invariant system that
is plagued by cancellations
between strong and Coulomb interactions.
Of course, fine-tuning has long been discussed in nuclear physics, but usually
in connection to the position of the Hoyle state of $^{12}$C
(see, for example, Ref. \cite{3alpha}).
An immediate question is to which extent the latter arises from
the fine-tunings we discussed here.

Together with previous and ongoing work on the $n\alpha$ \cite{BHvK1,BHvK2}
and $p\alpha$ systems \cite{BRvK}, our results provide a framework for
the description of cluster states in nuclei using EFT,
which can be seen as a generalization of the ERE to
systems that involve more than two bodies.
An important extension of this work is 
to systems of more than 
two alpha particles. 
The strong $\alpha\alpha$ interaction obtained
here, together with an exact treatment of the Coulomb
interaction, should provide the necessary ingredients.
Since $k_C/k_R\sim 3$ as discussed above, 
one might be able to simplify the calculation of Coulomb effects
by
developing a perturbation scheme in powers of 
$k_R/k_C$
from the start. 
Moreover, we conjecture that the $^{12}$C Hoyle state is 
a remnant of an Efimov state that appears in the unitary limit.
More complex $\alpha$-cluster systems could also be studied with
the EFT presented here.

\section*{Acknowledgments}
We would like to thank Carlos Bertulani for useful discussions
and encouragement.
RH has also benefited from discussions with Andreas Nogga, and
UvK with Brett Carlson, Tobias Frederico, Boris Gelman, and Manuel Malheiro.
This research was supported in part by the
Bundesministerium f\"ur Bildung und Forschung
under contract number 06BN411 (RH, HWH),
by the U.S. Department of Energy (UvK),
by the Nederlandse Organisatie voor Wetenschappelijk Onderzoek (UvK),
and by Brazil's FAPESP under a Visiting Professor grant (UvK).
UvK would like to thank the hospitality of 
the Kernfysisch Versneller Instituut at Rijksuniversiteit Groningen,
the Instituto de F\'\i sica Te\'orica of the
Universidade Estadual Paulista,
and 
the Instituto de F\'\i sica of the Universidade de S\~ao Paulo,
where part of this work was carried out.

\appendix
\section{Coulomb Green's function and EFT}
\label{app:CEFT}

The scattering amplitude for two particles in their center-of-mass
(CM) system interacting via Coulomb 
and a short-range interaction is given by \cite{GW}
\begin{equation}
T=T_C+T_{CS}
=\langle \chi^{(-)}_{k'}\vert\,V_{C}\,\vert {\mbox{\boldmath $k$}}\rangle
+\langle \chi^{(-)}_{k'}\vert\, V_{S}\,\vert \Psi^{(+)}_{k}\rangle
\,,
\end{equation}
where $|\mbox{\boldmath $k$}\rangle$, $|\chi^{(+/-)}_k\rangle$,
and $|\Psi^{(+/-)}_k\rangle$ represent free and 
(incoming/outgoing) states of momentum $\mbox{\boldmath $k$}$ 
for pure-Coulomb and Coulomb-distorted 
short-range interactions, respectively, while $V_C$ ($V_S$) is 
the Coulomb (short-range) interaction operator. In coordinate space, the 
Coulomb wave functions can be written as \cite{KR00}
\begin{equation}
\langle {\mbox{\boldmath $r$}}\vert \chi^{(\pm)}_{k}\rangle \equiv 
\chi^{(\pm)}_{k}({\mbox{\boldmath $r$}})=e^{-\frac{\eta\pi}{2}}\,
\Gamma(1\pm i\eta)\,M(\mp i\eta,1;\pm ikr-i{\mbox{\boldmath $k$}}\cdot 
{\mbox{\boldmath $r$}})\,
e^{i{\mbox{\scriptsize\boldmath $k$}}\cdot {\mbox{\scriptsize\boldmath $r$}}}
\,,
\label{eq:onlyCwf}
\end{equation}
where $M(a,b;z)$ is the Kummer function. From  
$M(a,b;0)=1$ \cite{abramo}, one obtains the important properties 
\begin{eqnarray}
&&\chi^{(\pm)\,*}_{k'}(0)\,\chi^{(\pm)}_{k}(0)=e^{-\pi\eta}\Gamma(1\mp i\eta)
\Gamma(1\pm i\eta)=\frac{2\pi\eta}{e^{2\pi\eta}-1}\equiv C_{\eta}^2\,,
\label{eq:psi0-1}
\\[1mm]
&&\chi^{(\mp)\,*}_{k'}(0)\,\chi^{(\pm)}_{k}(0)=e^{-\pi\eta}\Gamma(1\pm i\eta)^2
=C_{\eta}^2\,e^{\pm 2i\sigma_0}\,,
\label{eq:psi0-2}
\end{eqnarray}
where $\sigma_0$ is the Coulomb phase shift for the partial wave $l=0$.
The general expression for the Coulomb phase shift,
\begin{equation}
\sigma_l=\arg\Gamma(l+1+i\eta)=\frac{1}{2i}\ln\left[
\frac{\Gamma(l+1+i\eta)}{\Gamma(l+1-i\eta)}\right]\,,
\label{eq:sigma_l}
\end{equation}
is defined from the partial-wave expansion of the pure-Coulomb amplitude 
\begin{equation}
T_C=-\frac{2\pi}{\mu}\sum_{l=0}^{\infty}(2l+1)
\frac{\left(e^{2i\sigma_l}-1\right)}{2ik}\,P_l(\cos\theta)
=-\frac{2\pi}{\mu}\,f_C(\theta)\,.
\end{equation}
The explicit solution
\begin{equation}
f_C(\theta)=-\frac{\eta^2}{2k_C}\,\csc^2\theta/2\,\exp\left[
i\eta\ln(\csc^2\theta/2)+2i\sigma_0\right]
\end{equation}
leads to the well-known Mott scattering cross section, 
$\sigma_M=|f_C(\theta)+f_C(\pi-\theta)|^2$,
which holds at very low energies. 

The main ingredient in the calculation
of the Coulomb-distorted short-range amplitude is the 
Coulomb Green's function at energy $E$, 
\begin{equation}
G_C^{(\pm)}(E)=\frac{1}{E-H_C\pm i\epsilon}
=2\mu\int\frac{d^3q}{(2\pi)^3}
\frac{|\chi^{(\pm)}_q\rangle\langle\chi^{(\pm)}_q|}
{2\mu E-{\mbox{\boldmath $q$}}^2\pm i\epsilon}\,.
\end{equation}
Using the Lippmann-Schwinger equation, one is able to express 
$|\Psi^{(\pm)}_k\rangle$ in terms of multiple insertions of the operator 
$G_C^{(\pm)}V_S$ acting on the Coulomb states 
$|\chi^{(\pm)}_k\rangle$ \cite{KR00}, allowing the 
amplitude to be 
written as the sum 
\begin{equation}
\label{eq:tcs1}
T_{CS}=\sum_{n=0}^{\infty}\langle\chi^{(-)}_{k'}|V_S
\left(G_C^{(\pm)}V_S\right)^n|\chi^{(+)}_k\rangle\,.
\end{equation}
This amplitude can be decomposed into
partial waves, 
\begin{equation}
T_{CS}=\sum_{l=0}^{\infty}T_l\,P_l(\cos\theta)\,,
\qquad T_l=-\frac{2\pi}{\mu}\,(2l+1)\,
\frac{e^{2i\sigma_l}}{k(\cot\delta^c_l-i)}\,,
\end{equation}
in terms of the ``Coulomb-corrected'' phase shifts $\delta^c_l$. 

\begin{figure}[t]
\begin{center}
\includegraphics[height=3.3cm]{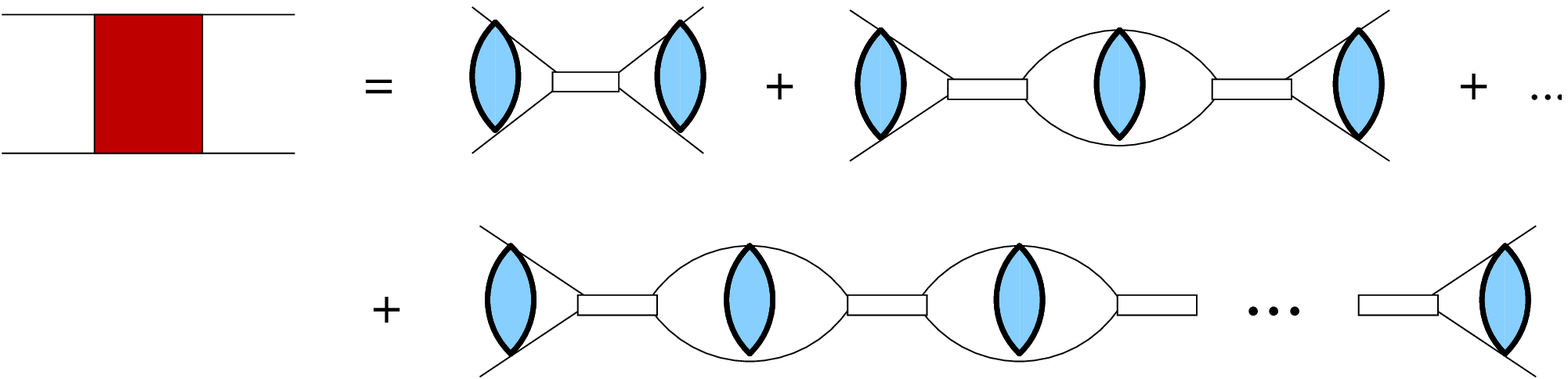}
\\[1cm]
\includegraphics[height=1.2cm]{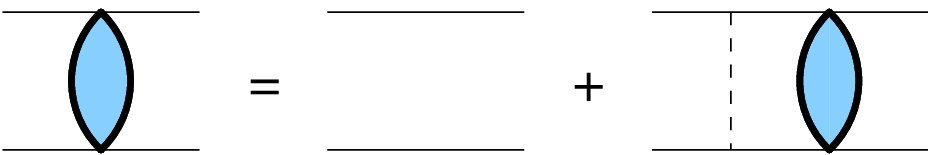}
\end{center}
\vspace*{-0pt}
\caption{Resummation for the LO expression of $T_{CS}$,
represented by the (red) shaded rectangle. 
The double line stands for the dimeron propagator.
The (blue) shaded ellipse represents the propagation
in the presence of Coulomb photons (dashed line) but
absence of a dimeron.
}
\label{fig:T0sum}
\end{figure}

If the deviation from the  Mott cross section
is mostly $S$ wave,
then the calculation of 
$T_{CS}$ proceeds from the Lagrangian (\ref{eq:LOLag}) 
similarly to Ref.~\cite{KR00}. 
The purely strong matrix element 
$\langle {\mbox{\boldmath $p'$}}|V_S|{\mbox{\boldmath $p$}}\rangle$ 
is defined as the amplitude $t_d=g^2 D_d(E;\mbox{\boldmath $0$})$ 
for $\alpha\alpha$ scattering via 
a bare dimeron two-point function, with the 
external legs amputated, evaluated in the CM: 
\begin{equation}
\langle {\mbox{\boldmath $p'$}}|V_S|{\mbox{\boldmath $p$}}\rangle 
=g^2 D_d(E;\mbox{\boldmath $0$})= \frac{\sigma\,g^2}{E-\Delta+i\epsilon}\,.
\label{eq:mtrx1}
\end{equation}
Explicit evaluation of Eq.~(\ref{eq:tcs1})
using Eqs.~(\ref{eq:psi0-1}), (\ref{eq:psi0-2}), 
and (\ref{eq:mtrx1}) then gives
\begin{equation}
\label{eq:tcs1_eval}
T_{CS}=\frac{\sigma\,g^2}{E-\Delta+i\epsilon}\,C_{\eta}^2\,e^{2i\sigma_0}
\left[1+\frac{\sigma\,g^2}{E-\Delta+i\epsilon}\,J_0(E)+\left(
\frac{\sigma\,g^2}{E-\Delta+i\epsilon}\,J_0(E)\right)^2+\cdots\right]\,.
\end{equation}
This resummation of the Coulomb exchanges 
is illustrated in  Fig.~\ref{fig:T0sum}.
It leads to
\begin{eqnarray}
T_{CS}&=&
-\frac{2\pi}{\mu}\,C_{\eta}^2\,e^{2i\sigma_0}\left[
\sigma\frac{2\pi\Delta}{\mu g^2}-\sigma\frac{2\pi E}{\mu g^2}-i\epsilon
+\frac{2\pi}{\mu}\,J_0(E)\right]^{-1}\,,
\end{eqnarray}
where $J_0$ is given by 
\begin{equation}
J_0(E)=-\frac{\mu}{2\pi}\left\{
\frac{\kappa}{D-3}
+2k_C\left[H(\eta)+\frac{1}{D-4}-\ln\left(
\frac{\kappa\sqrt{\pi}}{2k_C}\right)-1+\frac{3}{2}\,C_E\right]\right\}\,,
\label{eq:KR-s-13a}
\end{equation}
with $D$ the dimensionality of 
spacetime, $\kappa$ the renormalization scale, 
$k_C=Z_{\alpha}^2\alpha_{em}\mu=k\eta$, $C_E= 0.577...$
the Euler-Mascheroni constant,
and 
\begin{equation}
H(\eta)=\psi(i\eta)+\frac{1}{2i\eta}-\ln(i\eta)\,
\label{eq:Hdef}
\end{equation}
in terms of the digamma function $\psi$,
which obeys \cite{abramo}
\begin{equation}
\Re[\psi(i\eta)]=\Re[\psi(1+i\eta)]\,,
\qquad
\Im[\psi(i\eta)]=\frac{1}{2\eta}+\frac{\pi}{2}\,\coth\pi\eta\,.
\end{equation}

The dimeron mass gets renormalized by the non-perturbative 
free-particle and Coulomb loops, 
\begin{equation}
\sigma\frac{2\pi\Delta^{(R)}}{\mu g^2}=
\sigma\frac{2\pi\Delta(\kappa)}{\mu g^2}
-\frac{\kappa}{D-3}
-2k_C\left[\frac{1}{D-4}
-\ln\left(\frac{\kappa\sqrt{\pi}}{2k_C}\right)-1+\frac{3}{2}\,C_E\right]\,,
\end{equation}
and $T_{CS}$ finally becomes 
\begin{eqnarray}
T_{CS}=-\frac{2\pi}{\mu}\,C_{\eta}^2\,e^{2i\sigma_0}\left[
\sigma\frac{2\pi\Delta^{(R)}}{\mu g^2}-\sigma\frac{\pi}{\mu^2 g^2}\,
k^2-2k_C H(\eta)\right]^{-1}\,.
\end{eqnarray}

\end{document}